\documentclass[12pt]{article}
\usepackage[utf8]{inputenc}
\usepackage{authblk}
\usepackage{setspace}
\usepackage[margin=1.25in]{geometry}
\usepackage{graphicx}
\usepackage{subcaption}
\usepackage{amsmath}
\usepackage{lineno}
\usepackage{lipsum}
\usepackage[style=nejm, 
citestyle=numeric-comp,
sorting=none]{biblatex}
\begin{document}
\title{Probing nuclear structure using elliptic flow of strange and multi-strange hadrons in isobar collisions}
\author[1*]{Priyanshi Sinha (for the STAR collaboration)}
\affil[1]{Indian Institute of Science Education and Research (IISER) Tirupati}
\affil[*]{priyanshisinha@students.iisertirupati.ac.in}

\onehalfspacing
\maketitle

\date{}

%%%%%% Abstract %%%%%%
\begin{abstract}
Isobar collisions, $^{96}_{44}$Ru+$^{96}_{44}$Ru and $^{96}_{40}$Zr+$^{96}_{40}$Zr, at $\sqrt{s_{\mathrm {NN}}}$ = 200 GeV have been performed at RHIC in order to study the charge separation along the magnetic field, called the Chiral Magnetic Effect (CME). The difference in nuclear deformation and structure between the two isobar nuclei may result in a difference in the flow magnitudes. Hence, elliptic flow measurements for these collisions give direct information about the initial state anisotropies. Strange and multi-strange hadrons have a small hadronic cross-section compared to light hadrons, making them an excellent probe for understanding the initial state anisotropies of the medium produced in these isobar collisions. The collected datasets include approximately two billion events for each of the isobar species and provide a unique opportunity for statistics hungry measurements. 
In this proceeding, we will report the elliptic flow ($v_{2}$) measurement of $K_{s}^{0}$, $\Lambda$, $\overline{\Lambda}$, $\phi$, $\Xi^{-}$,  $\overline{\Xi}^{+}$, $\Omega^{-}$, and $\overline{\Omega}^{+}$ at mid-rapidity for Ru+Ru and Zr+Zr collisions at $\sqrt{s_{\mathrm {NN}}}$ = 200 GeV. The centrality and transverse momentum ($p_{T}$) dependence of $v_{2}$ of (multi-)strange hadrons will be shown. System size dependence of $v_{2}$ will be shown by comparing the $v_{2}$ results obtained from Cu+Cu, Au+Au, and U+U collisions. The number of constituent quark (NCQ) scaling for these strange hadrons will also be tested. We will also compare the $p_{T}$-integrated $v_{2}$ for these two isobar collisions. Transport model calculations will be compared to data to provide further quantitative constraints on the nuclear structure.
\end{abstract}
\newpage
\section{Introduction}
Relativistic heavy-ion collisions indicate the presence of a strongly interacting medium called Quark Gluon Plasma (QGP).  Studies of the elliptic flow of produced particles in this medium provide insight into the early anisotropy in the medium. Despite having the same number of nucleons, the anisotropic flow coefficients of Ru+Ru and Zr+Zr were observed to be distinct in the isobar collision run at $\sqrt{s_{\mathrm {NN}}}$ = 200 GeV at the Relativistic Heavy Ion Collider (RHIC)~\cite{isobarData}. This suggests that the difference in nuclear structure may also leave imprints on the elliptic flow of particles. Recent studies also discuss the use of $v_2$ ratios and $v_2$-[$p_T$] correlations in isobar collisions to probe nuclear structures.~\cite{betaDeform, v2ptcorrelation}. Compared to light hadrons, (multi-)strange hadrons have a smaller hadronic cross-section, making their $v_2$ an excellent probe of the initial state anisotropies in these isobar collisions.
\section{Analysis details}
We report strange and multi-strange hadron $v_{2}$ in $^{96}_{44}$Ru+$^{96}_{44}$Ru and $^{96}_{40}$Zr+$^{96}_{40}$Zr collisions at $\sqrt{s_{\mathrm {NN}}}$ = 200 GeV using the data collected by the STAR experiment.  Each isobar collision had roughly 650M events analysed. The Time Projection Chamber (TPC) and Time-Of-Flight (TOF) have been used to identify the decay daughters of these short-lived particles and reconstruct them using invariant mass technique. Using their weak-decay topology, strange particles $K_{s}^{0}$, $\Lambda(\bar{\Lambda})$, and $\Xi^{-}(\overline{\Xi}^{+})$ are reconstructed. The combinatorial background for these hadrons is constructed using rotational background method~\cite{multiStrange}. $\phi$-mesons are reconstructed using hadronic decay channel and event mixing technique is used for combinatorial background estimate. The $v_{2}$ is calculated using $\eta$-sub event plane method~\cite{Strange}. The maximum resolution for the second order event plane is nearly 48\% for both collision systems.
\section{Results}
\begin{figure}[h!]
%%\vspace{-.2cm}
\centering
\begin{tabular}{cccc}
 \includegraphics[width=3.5cm]{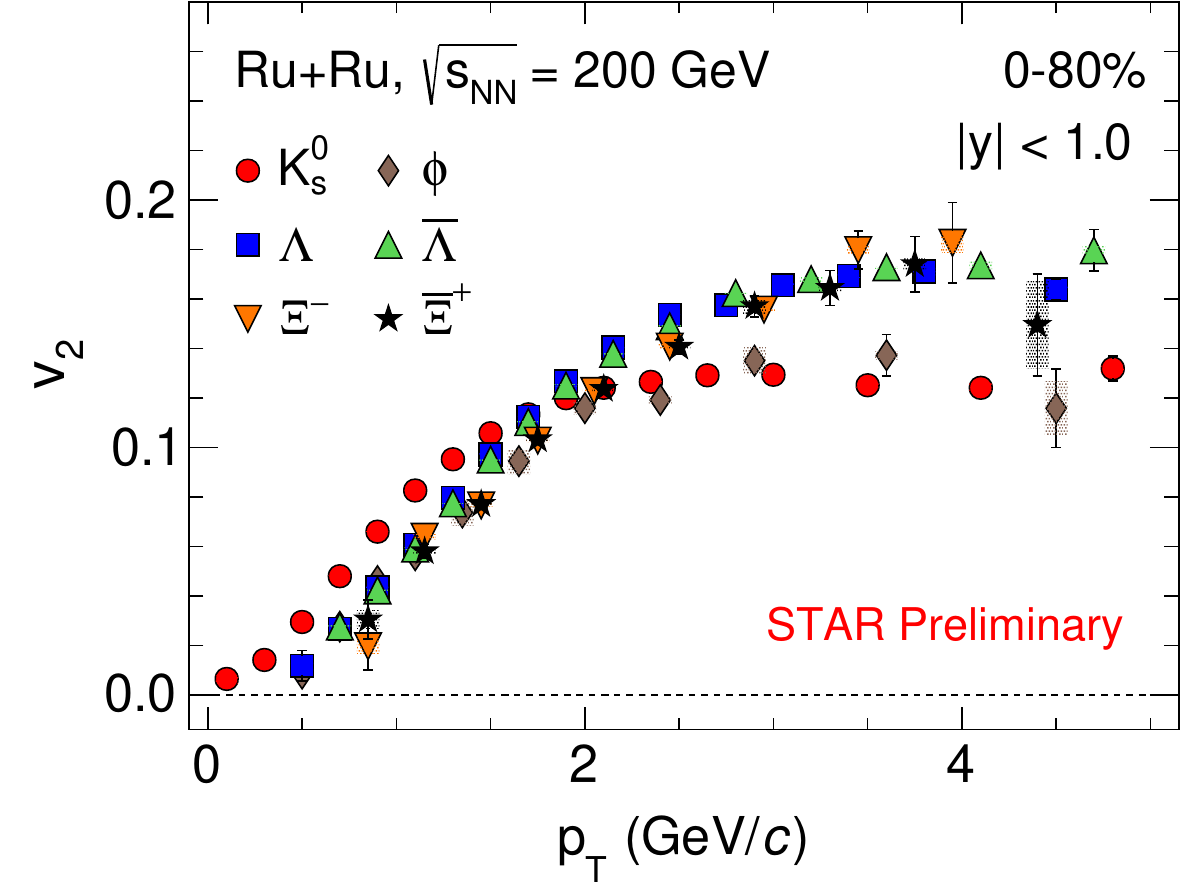} &
 \includegraphics[width=3.5cm]{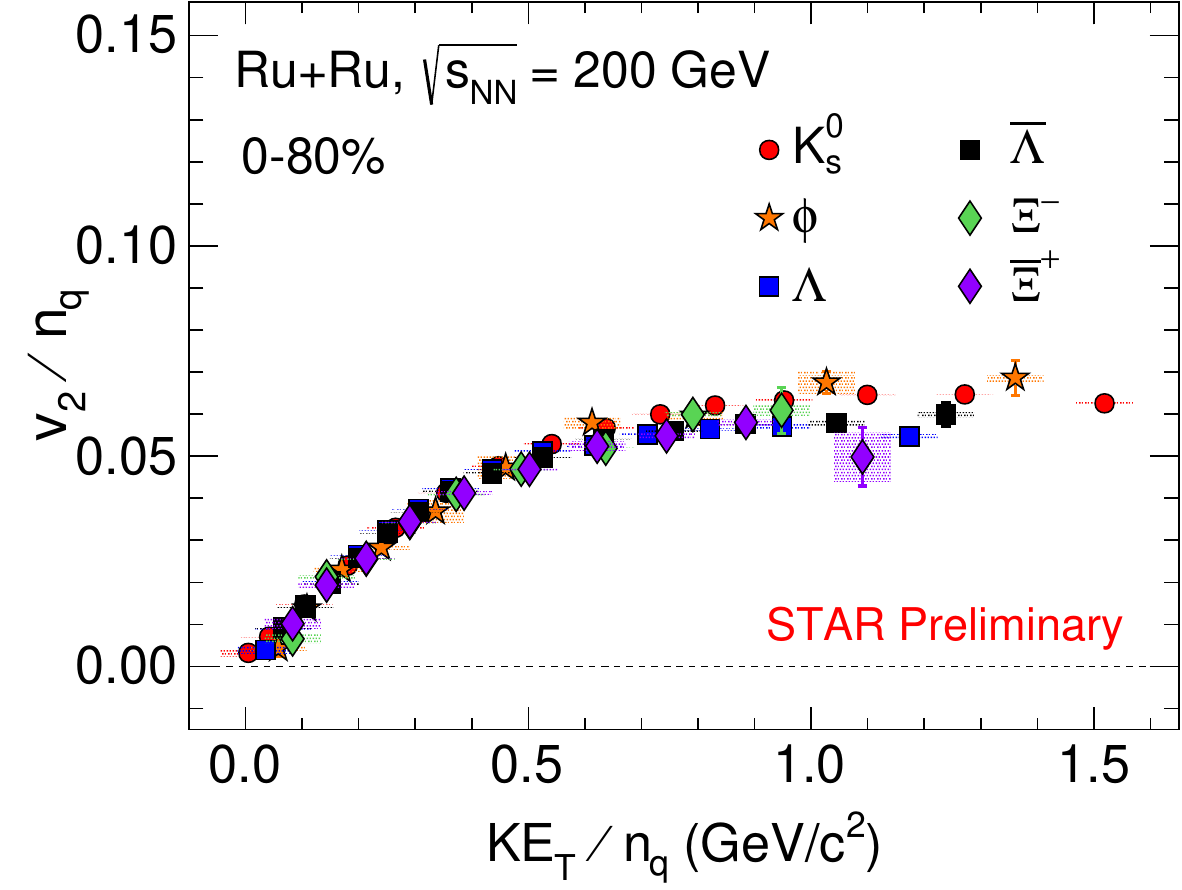} &
 \includegraphics[width=3.5cm]{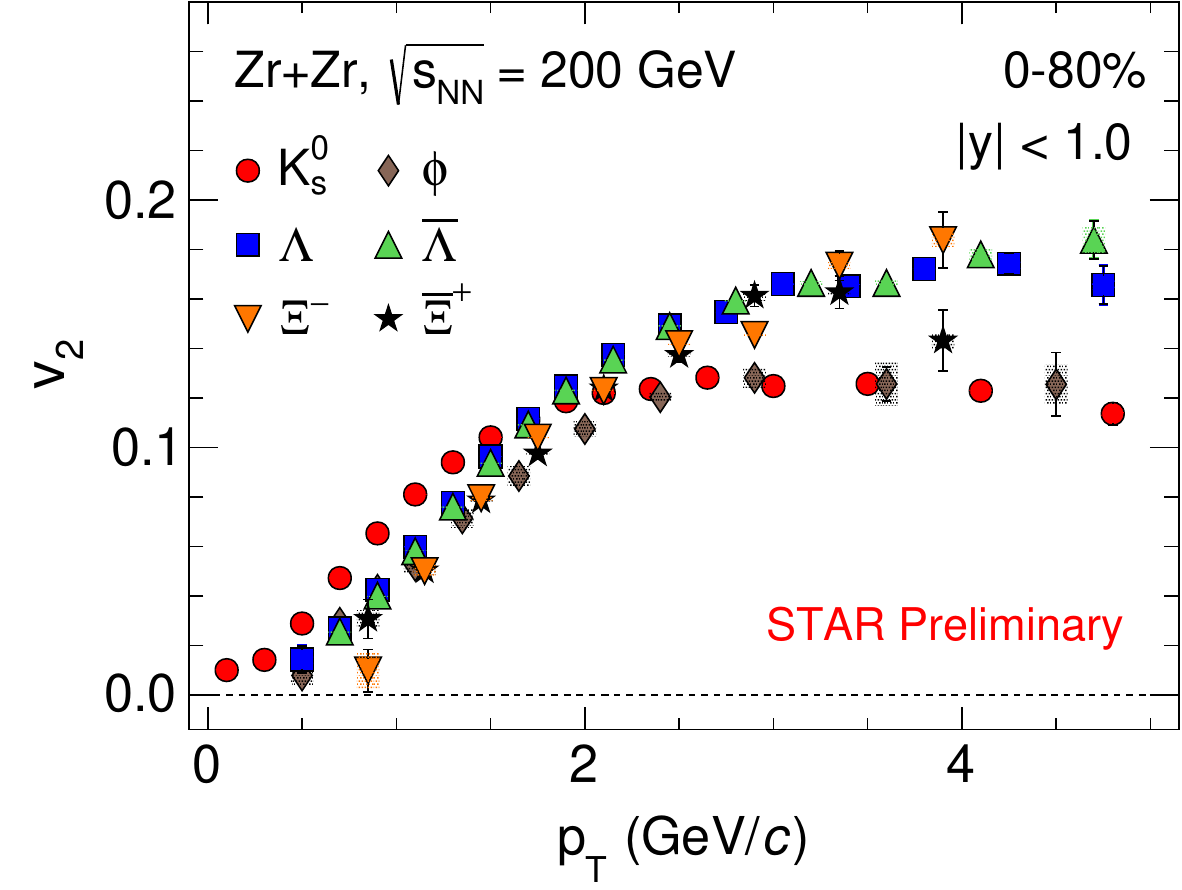} &
 \includegraphics[width=3.5cm]{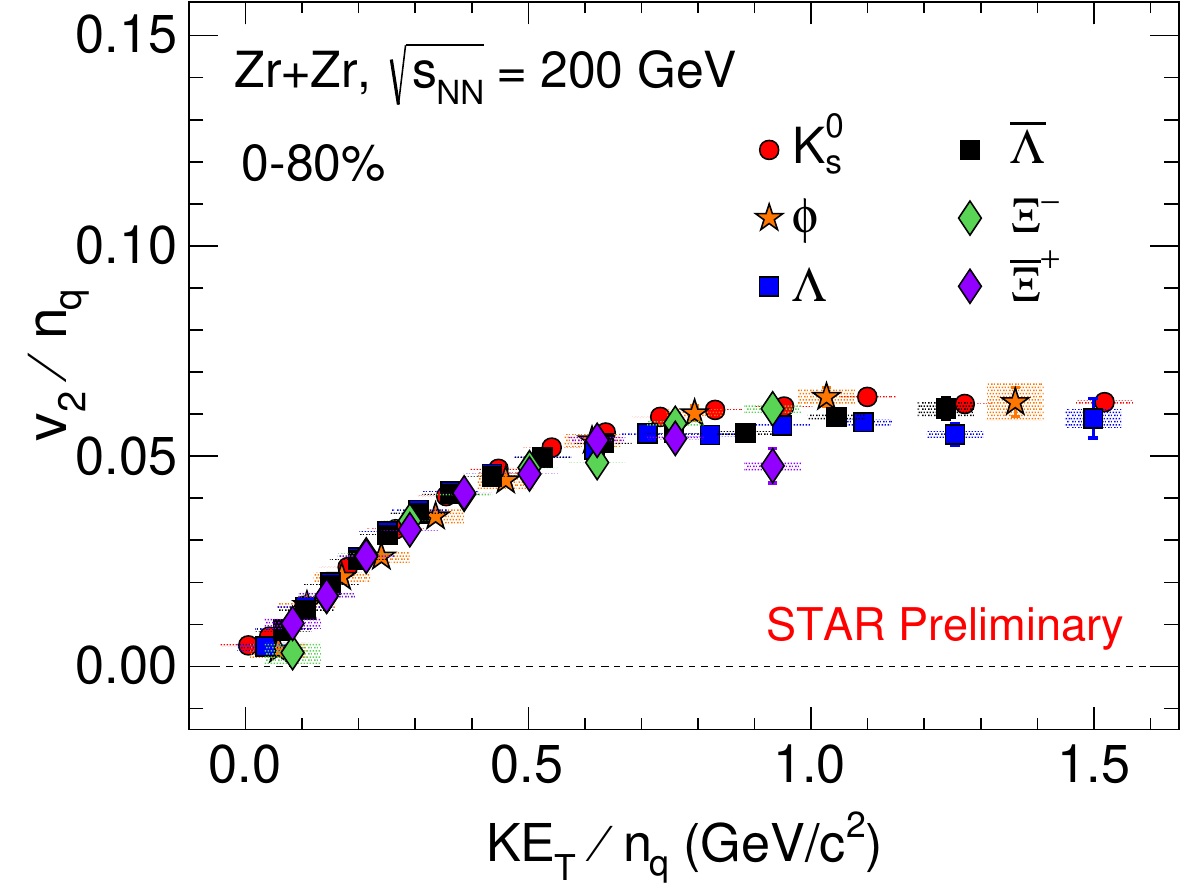}
\end{tabular}
%%\vspace{-0.2cm}
\caption{$v_{2}$ as a function of $p_{T}$ of (multi-)strange hadrons and NCQ-scaled $v_{2}$ as a function of transverse kinetic energy for Ru+Ru and Zr+Zr collisions at $\sqrt{s_{\mathrm {NN}}}$ = 200 GeV. The vertical lines and shaded boxes denote statistical and systematic uncertainties, respectively.}
%%\vspace{-0.3cm}
\label{fig:Fig.1}
\end{figure}

\begin{figure}[h!]
\centering
\begin{tabular}{ccc}
\includegraphics[width=3.7cm]{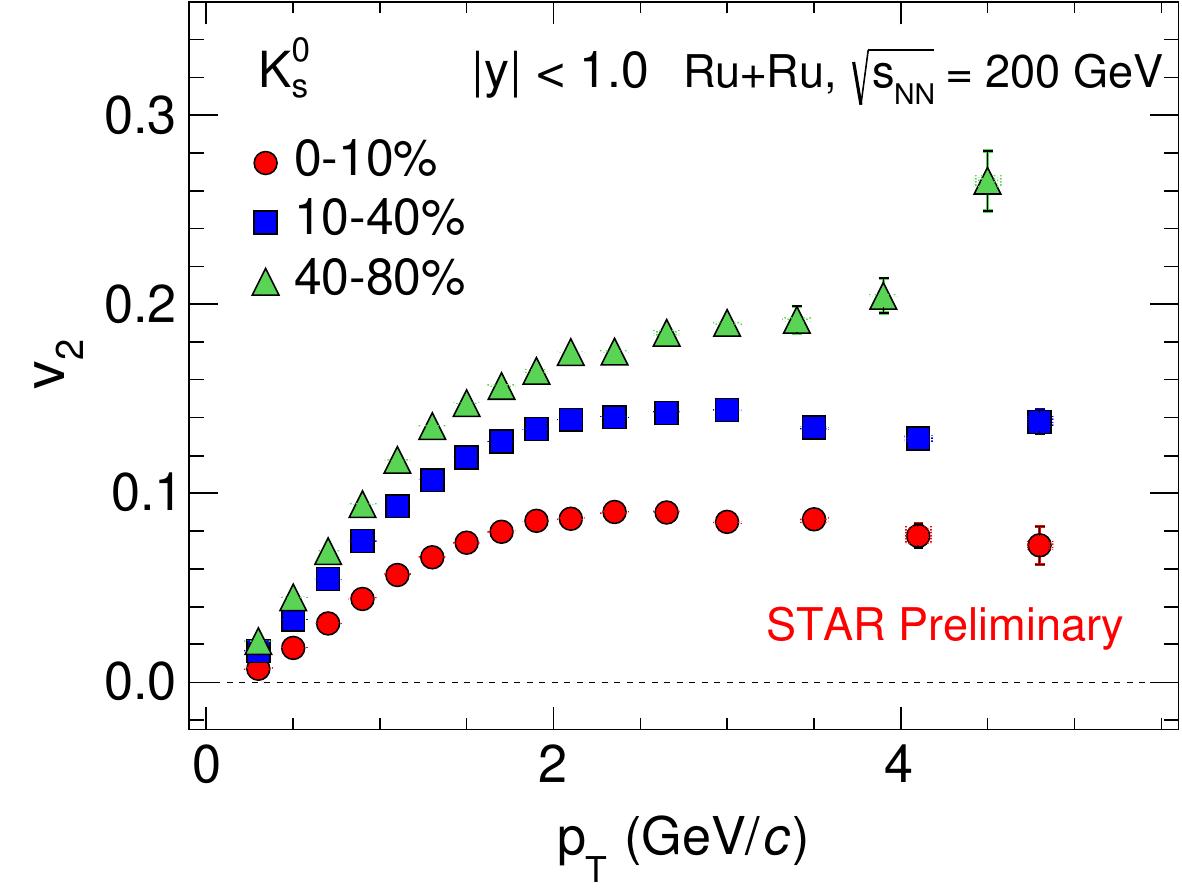} &
 \includegraphics[width=3.7cm]{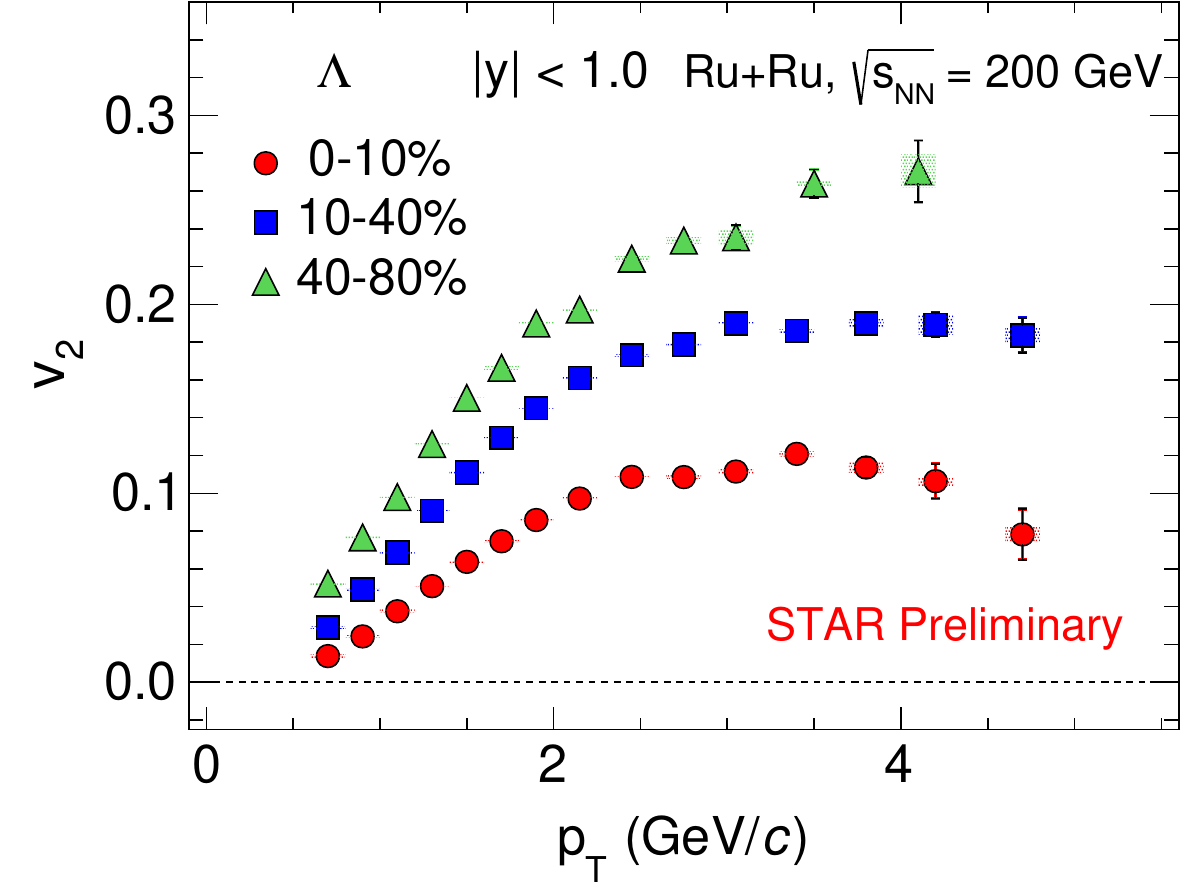} &
  \includegraphics[width=3.7cm]{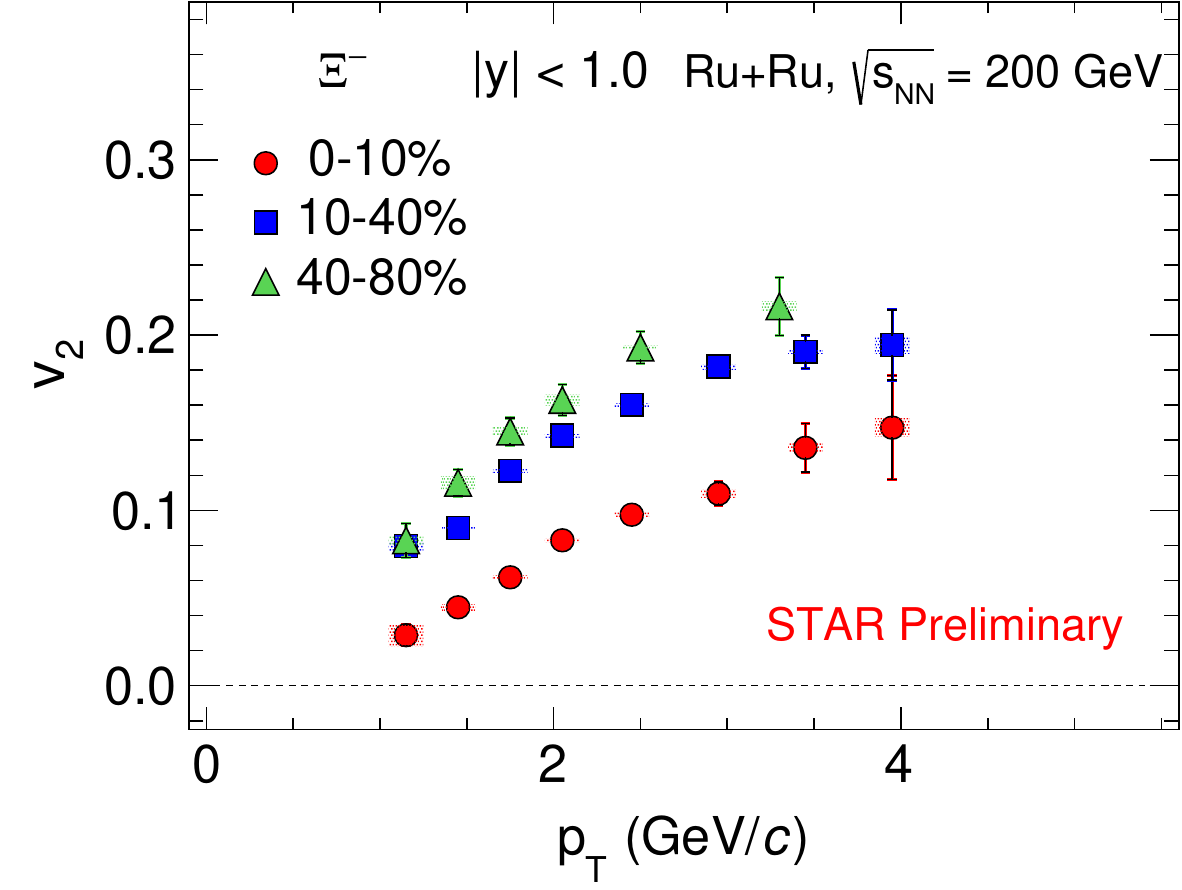} \\
 \includegraphics[width=3.7cm]{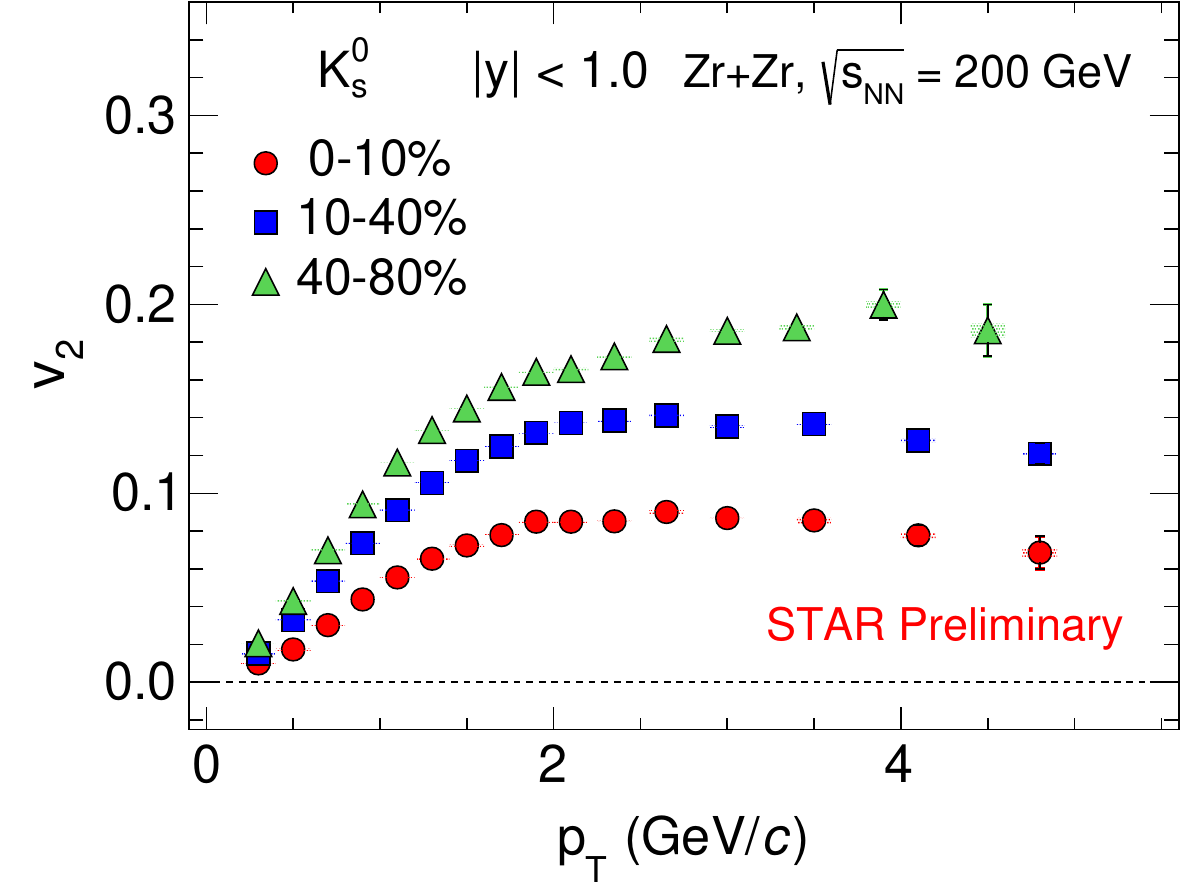}&
 \includegraphics[width=3.7cm]
 {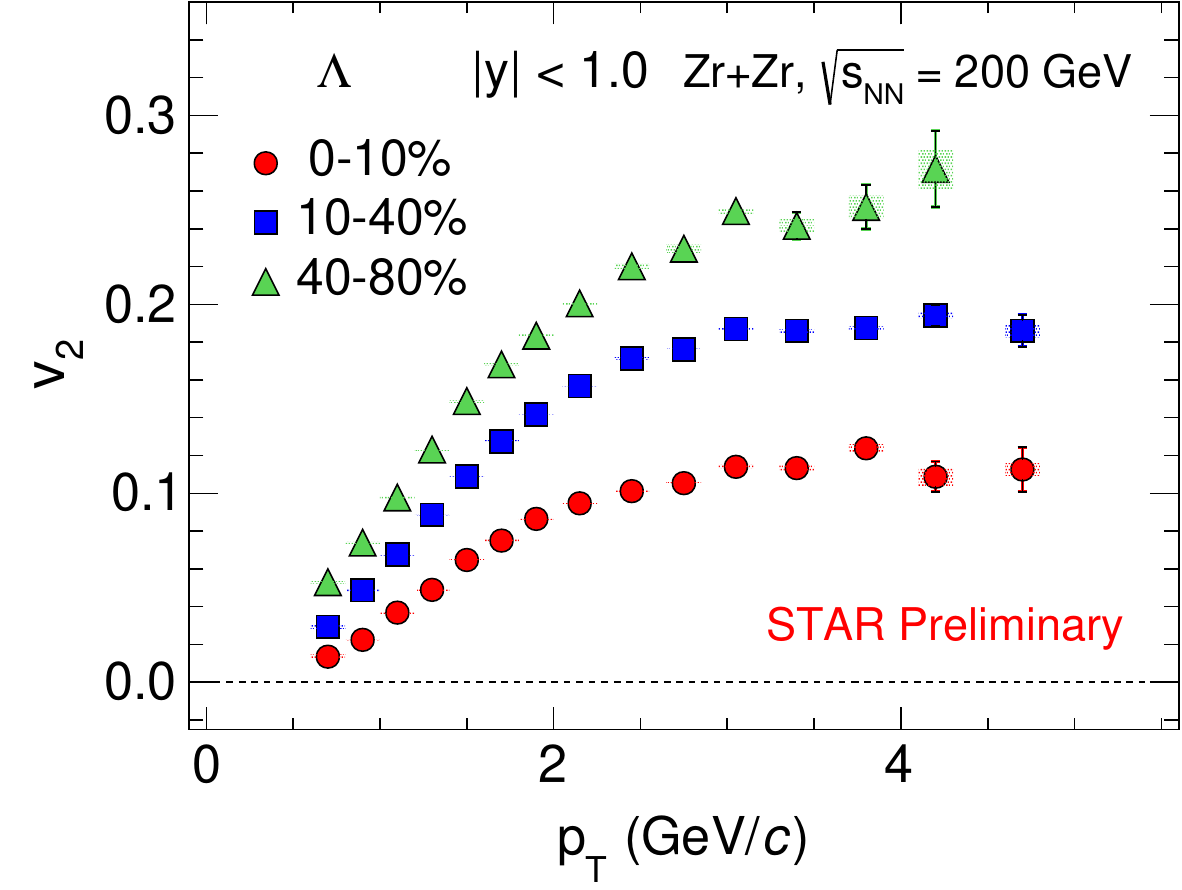} &
 \includegraphics[width=3.7cm]{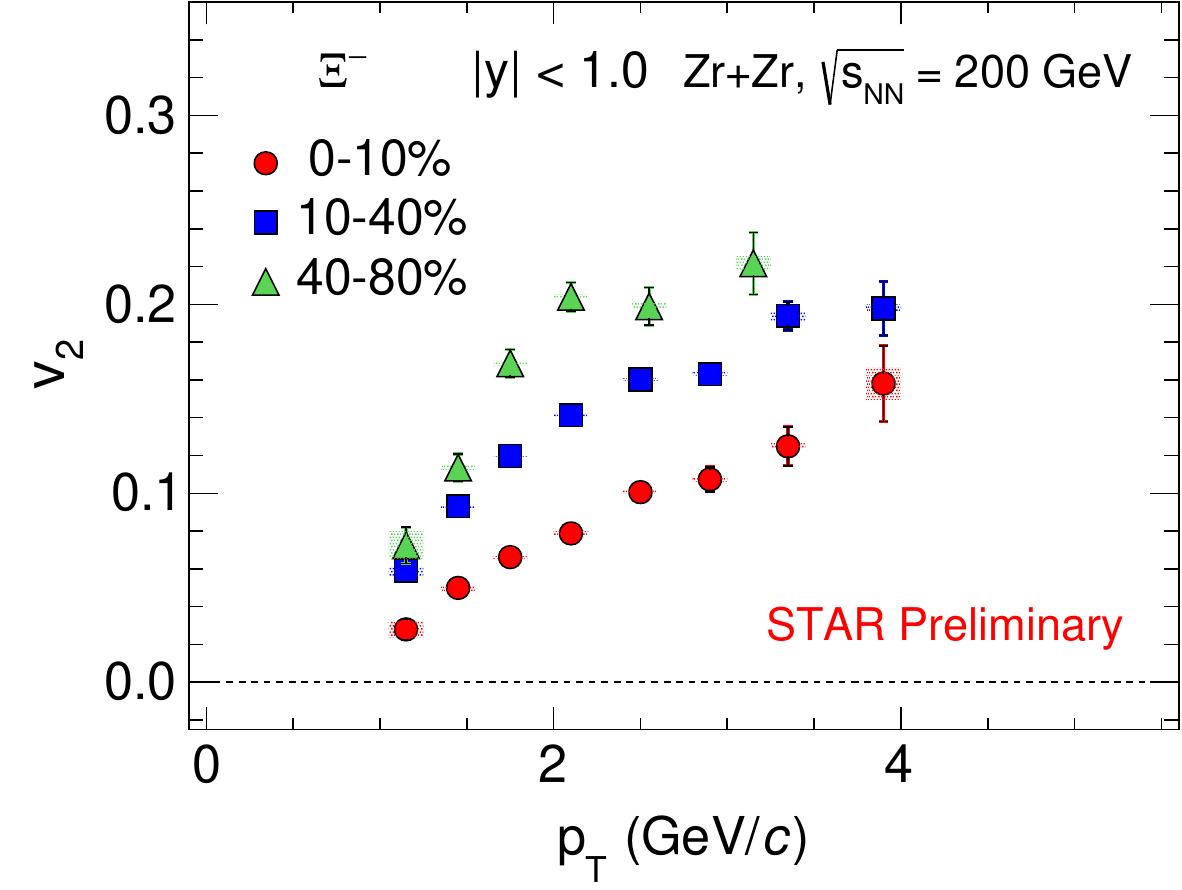}\\ 
\end{tabular}
%%\vspace{-0.2cm}
\caption{Top panel: Centrality dependence of $v_{2}$ of $K_{s}^{0}$, $\Lambda$ and $\Xi^{-}$ as a function of $p_{T}$ in Ru+Ru collisions at $\sqrt{s_{\mathrm {NN}}}$ = 200 GeV; Bottom Panel: Same for Zr+Zr collisions at $\sqrt{s_{\mathrm {NN}}}$ = 200 GeV. The vertical lines and shaded boxes denote statistical and systematic uncertainties, respectively.}
%\vspace{-0.2cm}
\label{fig:Fig.2}
\end{figure}

\begin{figure}[h!]
\centering
 \includegraphics[width=11cm]{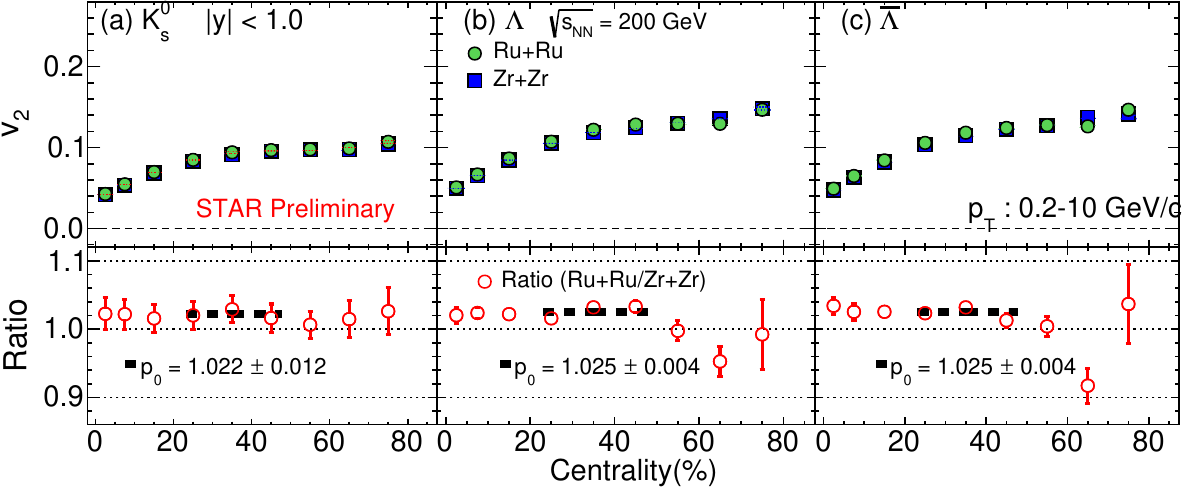} 
 %%\vspace{-0.25cm}
\caption{$p_{T}$-integrated $v_{2}$ as a function of centrality for $K_{s}^{0}$, $\Lambda$, and $\bar{\Lambda}$ in Ru+Ru and Zr+Zr collisions at $\sqrt{s_{\mathrm {NN}}}$ = 200 GeV. The vertical lines on the ratio includes statistical and systematic uncertainties. The dotted lines denotes the fitting with a constant.}
%\vspace{-0.2cm}
\label{fig:Fig.3}
\end{figure}

Figure ~\ref{fig:Fig.1} shows the $v_{2}$ of strange and multi-strange hadrons as a function of $p_{T}$ for minimum bias Ru+Ru and Zr+Zr collisions at $\sqrt{s_{\mathrm {NN}}}$ = 200 GeV. An approximate mass ordering at low $p_{T}$ and a baryon-meson splitting at intermediate $p_{T}$ is observed. All particles and anti-particles tend to follow the number of constituent quark (NCQ) scaling within 10\% as shown in Fig.~\ref{fig:Fig.1}, indicating partonic collectivity as well as domination of quark coalescence mechanism for hadronization at intermediate $p_{T}$ region. 
A clear centrality dependence of $v_{2}$ has been observed for $K_{s}^{0}$, $\Lambda$ and $\Xi^{-}$ as shown in Fig.~\ref{fig:Fig.2} and for other hadrons for the isobar collision systems. 
The $p_{T}$-integrated $v_{2}$ for strange hadrons was also studied as a function of the collision centrality as shown in Fig.~\ref{fig:Fig.3}. The ratios of $v_{2}$ between the two isobar collisions for $K_{s}^{0}$, $\Lambda$, and $\bar{\Lambda}$ show clear deviation of nearly 2\% from unity in mid-central collisions, indicating a difference in nuclear structure and shape~\cite{isobarData}.\\
\begin{figure}[h!]
\centering
 \includegraphics[width=11cm]{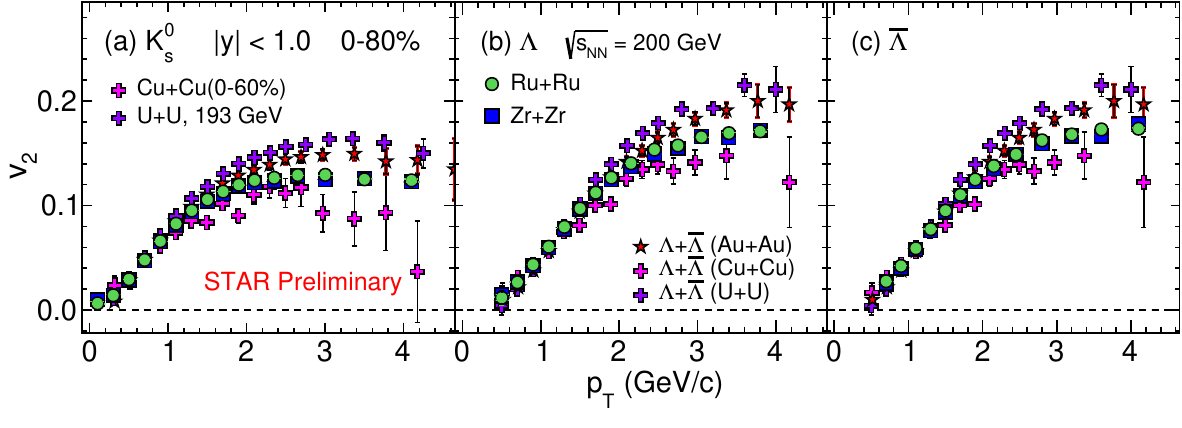} 
% %\vspace{-0.4cm}
\caption{$v_{2}$ of $K_{s}^{0}$, $\Lambda$ and $\bar{\Lambda}$ in minimum bias Cu+Cu, Ru+Ru, Zr+Zr, Au+Au collisions at $\sqrt{s_{\mathrm {NN}}}$ = 200 GeV and U+U collisions at $\sqrt{s_{\mathrm {NN}}}$ = 193 GeV~\cite{CuData, AuData, UData}.}
%%\vspace{-0.5cm}
\label{fig:Fig.4}
\end{figure}
We investigated the system size evolution of $v_{2}$ by comparing the $^{63}_{29}$Cu+$^{63}_{29}$Cu, $^{96}_{44}$Ru+$^{96}_{44}$Ru, $^{96}_{40}$Zr+$^{96}_{40}$Zr, $^{197}_{79}$Au+$^{197}_{79}$Au collisions in 0-80\% centrality at $\sqrt{s_{\mathrm {NN}}}$ = 200 GeV, and $^{238}_{92}$U+$^{238}_{92}$U collisions at $\sqrt{s_{\mathrm {NN}}}$ = 193 GeV~\cite{CuData, AuData, UData}. Figure \ref{fig:Fig.4} shows an approximate system size dependence of $v_{2}$ for $p_{T}$ $>$ 1.8 GeV/$c$, based on the nuclear size. The  $v_{2}$ in U+U and Au+Au is observed to be higher, whereas in Cu+Cu is slightly lower than those in isobar collisions. 
\section{Conclusion}
%%\vspace{-0.5cm}
We presented the elliptic flow of $K_{s}^{0}$, $\Lambda$, $\bar{\Lambda}$, $\phi$, $\Xi^{-}$, and $\overline{\Xi}^{+}$ particles in Ru+Ru and Zr+Zr collisions at $\sqrt{s_{\mathrm {NN}}}$ = 200 GeV. In these isobar collisions, we noticed a mass ordering at low $p_{T}$ and a baryon-meson splitting at intermediate $p_{T}$. The NCQ scaling representing the partonic degrees of freedom and coalescence hadronization, is followed by all strange hadrons. The hadron $v_2$ ratio exhibits a deviation from unity of around 2\% when integrated over $p_{T}$ indicative of different nuclear density and deformation between the two isobar nuclei. The $v_{2}$ is observed to be higher for larger colliding systems. These results when compared to model calculations may shed light on structures of these nuclei.
%\lipsum[3]

\printbibliography

@article{isobarData,
   title = {Search for the chiral magnetic effect with isobar collisions at $\sqrt{{s}_{NN}}=200$ GeV by the STAR Collaboration at the BNL Relativistic Heavy Ion Collider},
   author = {Abdallah, M. S. and Aboona, B. E. and Adam, J. and Adamczyk, L. and Adams, J. R. and Adkins, J. K. and Agakishiev, G. and Aggarwal, I. and Aggarwal, M. M. and Ahammed, Z. and Alekseev, I. and Anderson, D. M. and Aparin, A. and Aschenauer, E. C. and Ashraf, M. U. and Atetalla, F. G. and Attri, A. and Averichev, G. S. and Bairathi, V. and Baker, W. and Ball Cap, J. G. and Barish, K. and Behera, A. and Bellwied, R. and Bhagat, P. and Bhasin, A. and Bielcik, J. and Bielcikova, J. and Bordyuzhin, I. G. and Brandenburg, J. D. and Brandin, A. V. and Bunzarov, I. and Cai, X. Z. and Caines, H. and Calder\'on de la Barca S\'anchez, M. and Cebra, D. and Chakaberia, I. and Chaloupka, P. and Chan, B. K. and Chang, F-H. and Chang, Z. and Chankova-Bunzarova, N. and Chatterjee, A. and Chattopadhyay, S. and Chen, D. and Chen, J. and Chen, J. H. and Chen, X. and Chen, Z. and Cheng, J. and Chevalier, M. and Choudhury, S. and Christie, W. and Chu, X. and Crawford, H. J. and Csan\'ad, M. and Daugherity, M. and Dedovich, T. G. and Deppner, I. M. and Derevschikov, A. A. and Dhamija, A. and Di Carlo, L. and Didenko, L. and Dixit, P. and Dong, X. and Drachenberg, J. L. and Duckworth, E. and Dunlop, J. C. and Elsey, N. and Engelage, J. and Eppley, G. and Esumi, S. and Evdokimov, O. and Ewigleben, A. and Eyser, O. and Fatemi, R. and Fawzi, F. M. and Fazio, S. and Federic, P. and Fedorisin, J. and Feng, C. J. and Feng, Y. and Filip, P. and Finch, E. and Fisyak, Y. and Francisco, A. and Fu, C. and Fulek, L. and Gagliardi, C. A. and Galatyuk, T. and Geurts, F. and Ghimire, N. and Gibson, A. and Gopal, K. and Gou, X. and Grosnick, D. and Gupta, A. and Guryn, W. and Hamad, A. I. and Hamed, A. and Han, Y. and Harabasz, S. and Harasty, M. D. and Harris, J. W. and Harrison, H. and He, S. and He, W. and He, X. H. and He, Y. and Heppelmann, S. and Heppelmann, S. and Herrmann, N. and Hoffman, E. and Holub, L. and Hu, Y. and Huang, H. and Huang, H. Z. and Huang, S. L. and Huang, T. and Huang, X. and Huang, Y. and Humanic, T. J. and Igo, G. and Isenhower, D. and Jacobs, W. W. and Jena, C. and Jentsch, A. and Ji, Y. and Jia, J. and Jiang, K. and Ju, X. and Judd, E. G. and Kabana, S. and Kabir, M. L. and Kagamaster, S. and Kalinkin, D. and Kang, K. and Kapukchyan, D. and Kauder, K. and Ke, H. W. and Keane, D. and Kechechyan, A. and Kelsey, M. and Khyzhniak, Y. V. and Kiko\l{}a, D. P. and Kim, C. and Kimelman, B. and Kincses, D. and Kisel, I. and Kiselev, A. and Knospe, A. G. and Ko, H. S. and Kochenda, L. and Kosarzewski, L. K. and Kramarik, L. and Kravtsov, P. and Kumar, L. and Kumar, S. and Kunnawalkam Elayavalli, R. and Kwasizur, J. H. and Lacey, R. and Lan, S. and Landgraf, J. M. and Lauret, J. and Lebedev, A. and Lednicky, R. and Lee, J. H. and Leung, Y. H. and Li, C. and Li, C. and Li, W. and Li, X. and Li, Y. and Li, Y. and Liang, X. and Liang, Y. and Licenik, R. and Lin, T. and Lin, Y. and Lisa, M. A. and Liu, F. and Liu, H. and Liu, H. and Liu, P. and Liu, T. and Liu, X. and Liu, Y. and Liu, Z. and Ljubicic, T. and Llope, W. J. and Longacre, R. S. and Loyd, E. and Lukow, N. S. and Luo, X. F. and Ma, L. and Ma, R. and Ma, Y. G. and Magdy, N. and Mallick, D. and Margetis, S. and Markert, C. and Matis, H. S. and Mazer, J. A. and Minaev, N. G. and Mioduszewski, S. and Mohanty, B. and Mondal, M. M. and Mooney, I. and Morozov, D. A. and Mukherjee, A. and Nagy, M. and Nam, J. D. and Nasim, Md. and Nayak, K. and Neff, D. and Nelson, J. M. and Nemes, D. B. and Nie, M. and Nigmatkulov, G. and Niida, T. and Nishitani, R. and Nogach, L. V. and Nonaka, T. and Nunes, A. S. and Odyniec, G. and Ogawa, A. and Oh, S. and Okorokov, V. A. and Page, B. S. and Pak, R. and Pan, J. and Pandav, A. and Pandey, A. K. and Panebratsev, Y. and Parfenov, P. and Pawlik, B. and Pawlowska, D. and Perkins, C. and Pinsky, L. and Pint\'er, R. L. and Pluta, J. and Pokhrel, B. R. and Ponimatkin, G. and Porter, J. and Posik, M. and Prozorova, V. and Pruthi, N. K. and Przybycien, M. and Putschke, J. and Qiu, H. and Quintero, A. and Racz, C. and Radhakrishnan, S. K. and Raha, N. and Ray, R. L. and Reed, R. and Ritter, H. G. and Robotkova, M. and Rogachevskiy, O. V. and Romero, J. L. and Roy, D. and Ruan, L. and Rusnak, J. and Sahoo, A. K. and Sahoo, N. R. and Sako, H. and Salur, S. and Sandweiss, J. and Sato, S. and Schmidke, W. B. and Schmitz, N. and Schweid, B. R. and Seck, F. and Seger, J. and Sergeeva, M. and Seto, R. and Seyboth, P. and Shah, N. and Shahaliev, E. and Shanmuganathan, P. V. and Shao, M. and Shao, T. and Sheikh, A. I. and Shen, D. Y. and Shi, S. S. and Shi, Y. and Shou, Q. Y. and Sichtermann, E. P. and Sikora, R. and Simko, M. and Singh, J. and Singha, S. and Skoby, M. J. and Smirnov, N. and S\"ohngen, Y. and Solyst, W. and Sorensen, P. and Spinka, H. M. and Srivastava, B. and Stanislaus, T. D. S. and Stefaniak, M. and Stewart, D. J. and Strikhanov, M. and Stringfellow, B. and Suaide, A. A. P. and Sumbera, M. and Summa, B. and Sun, X. M. and Sun, X. and Sun, Y. and Sun, Y. and Surrow, B. and Svirida, D. N. and Sweger, Z. W. and Szymanski, P. and Tang, A. H. and Tang, Z. and Taranenko, A. and Tarnowsky, T. and Thomas, J. H. and Timmins, A. R. and Tlusty, D. and Todoroki, T. and Tokarev, M. and Tomkiel, C. A. and Trentalange, S. and Tribble, R. E. and Tribedy, P. and Tripathy, S. K. and Truhlar, T. and Trzeciak, B. A. and Tsai, O. D. and Tu, Z. and Ullrich, T. and Underwood, D. G. and Upsal, I. and Van Buren, G. and Vanek, J. and Vasiliev, A. N. and Vassiliev, I. and Verkest, V. and Videb\ae{}k, F. and Vokal, S. and Voloshin, S. A. and Wang, F. and Wang, G. and Wang, J. S. and Wang, P. and Wang, Y. and Wang, Y. and Wang, Z. and Webb, J. C. and Weidenkaff, P. C. and Wen, L. and Westfall, G. D. and Wieman, H. and Wissink, S. W. and Wu, J. and Wu, J. and Wu, Y. and Xi, B. and Xiao, Z. G. and Xie, G. and Xie, W. and Xu, H. and Xu, N. and Xu, Q. H. and Xu, Y. and Xu, Z. and Xu, Z. and Yang, C. and Yang, Q. and Yang, S. and Yang, Y. and Ye, Z. and Ye, Z. and Yi, L. and Yip, K. and Yu, Y. and Zbroszczyk, H. and Zha, W. and Zhang, C. and Zhang, D. and Zhang, J. and Zhang, S. and Zhang, S. and Zhang, X. P. and Zhang, Y. and Zhang, Y. and Zhang, Y. and Zhang, Z. J. and Zhang, Z. and Zhang, Z. and Zhao, J. and Zhou, C. and Zhu, X. and Zurek, M. and Zyzak, M.},
collaboration = {STAR Collaboration},
  authortype = {(STAR Collaboration)},
   doi = {10.1103/PhysRevC.105.014901},
  journal = {Phys. Rev. C},
  volume = {105},
  pages = {014901},
  year = {2022}, 
}

@article{betaDeform,
     title = {Evidence of Quadrupole and Octupole Deformations in $^{96}\mathrm{Zr}+^{96}\mathrm{Zr}$ and $^{96}\mathrm{Ru}+^{96}\mathrm{Ru}$ Collisions at Ultrarelativistic Energies},
  author = {Zhang, Chunjian and Jia, Jiangyong},
  journal = {Phys. Rev. Lett.},
  volume = {128},
  pages = {022301},
  numpages = {6},
  year = {2022},
  month = {Jan},
  publisher = {American Physical Society},
  doi = {10.1103/PhysRevLett.128.022301},
  url = {https://link.aps.org/doi/10.1103/PhysRevLett.128.022301}
}

@article{v2ptcorrelation,
  title = {Probing nuclear quadrupole deformation from correlation of elliptic flow and transverse momentum in heavy ion collisions},
  author = {Jia, Jiangyong and Huang, Shengli and Zhang, Chunjian},
  journal = {Phys. Rev. C},
  volume = {105},
  pages = {014906},
  numpages = {12},
  year = {2022},
  month = {Jan},
  publisher = {American Physical Society},
  doi = {10.1103/PhysRevC.105.014906},
  url = {https://link.aps.org/doi/10.1103/PhysRevC.105.014906}
}

\end{document}